# Forecasts of the trend in global-mean temperature to 2100 arising from the scenarios of first-difference $CO_2$ and peak fossil fuel

L. Mark W. Leggett[1], David. A. Ball[1]

[1] Global Risk Policy Group Pty Ltd, Townsville, Queensland, Australia

*Correspondence to:* L. Mark W. Leggett (mleggett.globalriskprogress@gmail.com)

**Abstract.** Two future scenarios that are not explicitly in the range of scenarios (the Representative Concentration Pathway scenarios) utilised by the IPCC. These two scenarios are the emissions trend under peak fossil fuel (for example, Mohr et al., 2015); and the climate sensitivity determinable from the relationship between first-difference $CO_2$ and temperature recently shown by Leggett and Ball (2015). This paper provides forecasts of a global surface temperature trajectory to 2100 resulting from the effect of these two scenarios. The time-series models developed both displayed high statistical significance and converged in their forecasts, so adding to the potential robustness of the findings. Under the effect of the combination of the peak fossil fuel and first-difference $CO_2$ scenarios, we found that temperature is forecast to continue to rise, but only gently, until around 2023 where it reaches a level slightly higher than at present; and from then to display an also gentle steady decrease. It is shown that this trajectory is markedly lower than that generated by the IPCC business-as-usual level-of-CO2 (RCP8.5) model (Riahi et al. 2011). These lower results are evidence that the climate problem may require less future preventative action than is presently being considered necessary. If so, it is stressed that the same evidence is support for the case that the peak fossil fuel problem would require ameliorative action. This globally required action is the same as it would have been for climate (as embodied in the Paris Agreement on Climate Change of the 21st Conference of the Parties of the UNFCCC in Paris adopted in 2015) – the rapid transition to a predominantly renewable global energy system.

## 1 Introduction

Scenario planning (Lindgren and Bandhold 2003) has the goal of leading to a suite of policies for an entity which is robust against a range of alternative plausible futures.

At global level, one set of such futures concerns future global climate. One such future is the prospect of global climate worsening: this scenario has been termed by the Secretary-General of the United Nations, Ban Ki Moon, as no less than "the greatest collective challenge facing humankind today." (Ban 2014).

This worsening prospect arises because of the evidence for the following (IPCC 2013): increasing human carbon dioxide ($CO_2$) emissions leading to increasing atmospheric $CO_2$; this in turn leading to increasing global atmospheric temperature; and this in turn to an increasing frequency and intensity of other deleterious climate outcomes such as droughts, floods and storms, and other deleterious non-climate outcomes such as sea level rise.

The salience of this threat led to the creation in 1988 of the Intergovernmental Panel on Climate Change (IPCC). The IPCC (Weart 2011) is a scientific intergovernmental body set up under the auspices of the United Nations. The aim of the IPCC is to provide reports which provide UN governments with "the scientific, technical and socio-economic information relevant to understanding the scientific basis of risk of human-induced climate change, its potential impacts and options for adaptation and mitigation" (IPCC 1998).

In the latest such IPCC report, the Fifth Assessment Report of the Intergovernmental Panel on Climate Change (AR5) (IPCC 2013), the full range of plausible human emissions scenarios is distilled into four separate scenarios which span the range. These, termed Representative Concentration Pathways (RCPs), are derived for this century and beyond, and the future climate outcomes – such as global surface temperature – potentially arising are derived (van Vuuren et al. 2011) from the interaction of a range of different scenarios for society, the economy and policy. The scenarios range from one involving unfettered future emissions (RCP8.5) to three others involving various degrees of policy intervention or societal constraint.

The four RCP scenarios are at the foundation of the structure of the IPCC reports, and hence of the global intergovernmental process for addressing climate change.

With this background, this paper investigates the effect on the climate outcome of global surface temperature of a further two future scenarios that are not explicitly in the range of scenarios which led to the RCPs.

These two scenarios are the emissions trend under peak fossil fuel (for example, Mohr et al., 2015); and the climate sensitivity determinable from the relationship between first-difference $CO_2$ and global surface temperature recently shown by Leggett and Ball (2015).

### 1.1 Methodological considerations for forecasting

Finding the best method for quantitative forecasting involves several considerations: 1. What is the full range of models from which to choose? 2. Given that different models may have different strengths and weaknesses, what are the trade-offs to make such that an optimised modelling procedure can be chosen?

## 1.1.1 Spectrum of modelling types

Karplus (1977, 1992) has provided a framework for the characterisation and classification of models of systems. Enting (1987, 2010) has used this framework to assess model types used in climate studies.

In connection with his framework, Karplus (1992) observes that valid models of systems are the key to the successful prediction of the response (outputs) of systems to specified excitations (inputs). Karplus goes on to observe that there are numerous techniques, but all can be regarded as employing combinations of deduction and induction in varying proportions. *Deduction* means starting with something general and deriving something specific (we start with the law and deduce the model). *Induction* means starting with specific information and inferring something more general. In inductive modelling we gather and accumulate much data - specific values of inputs and outputs. When we think we have enough data, we try to arrive at a generalisation – a statement or perhaps an equation that applies to all or most of the data. Usually this entails discerning some sort of a pattern or a number of patterns in the data.

Karplus (1977) termed this deduction-induction range as a modelling spectrum. The position in the spectrum that each model took represented the degree of deduction as opposed to induction that is involved in the modelling process. The spectrum was described at the induction end as involving black-box models (characterised as being highly empirical, and only representing relations between inputs and outputs). The 'curve-fitting' model of a 'black-box' system is determined inductively from observations of the behaviour of the system. Curve-fitting, black-box models are typically used in fields such as economics.

At the other–deduction–end of the spectrum are white-box models (characterised as having relations between inputs and outputs defined through processes involving internal states of the system expressed in mechanistic terms). White-box models (Enting (2010) considers that 'glass-box models' would have been a better term) are generally deterministic. White-box models are frequently used in the physical sciences. The behaviour of a 'white-box' system can be deduced directly from a knowledge of the system structure and the basic physical laws that apply to it.

Climate studies have been assessed from the perspective of the Karplus spectrum by Enting (2010), who finds that both black-box and white-box models are used. Enting describes these as follows: **Black box:** Statistical fits which include regression-type analyses of $CO_2$ trends and cycles, correlation studies relating $CO_2$ and ENSO and empirical fits of transfer relations connecting concentrations to emissions. These apply at the globally aggregated level. Earth system: Earth system models, built around General Circulation Models (GCMs) for atmosphere and ocean represent the white-box end of the modelling spectrum for the earth system.

Enting (1987) argues that the spectrum concept gives a useful framework for comparing different types of modelling and forces an explicit recognition that possible uses of a model will depend on the type of model, i.e. its position within the spectrum.

Even though Enting shows the existence of both black-box and white-box (inductive/deductive) models in climate studies, the question arises as to the status of the usage of inductive/deductive *terminology and concepts* in the current IPCC assessment report (IPCC, 2013), and in the current climate literature?

In the IPCC Fifth Assessment Report (AR5) Glossary (IPCC 2013), despite the climate model entry being titled "Climate model (spectrum …)" the spectrum is limited to white-box types: there is no use of inductive/deductive, or black-box/white-box terminology. The definition reads: "A numerical representation of the climate system based on the physical, chemical and biological properties of its components, their interactions and feedback processes, and accounting for some of its known properties."

It can be seen that the climate model referred to here is synonymous with the AR5 Glossary definition for Process-based Model: "Theoretical concepts and computational methods that represent and simulate the behaviour of real-world systems derived from a set of functional components and their interactions with each other and the system environment, through physical and mechanistic processes occurring over time. "

Both of these descriptions clearly match with the white-box, deductive model type as described above.

None of the terms inductive, deductive, black-box, or white-box were found by computer search of the AR5 document.

That noted, the following statement from AR5 (page 825) is in strongly process-based/white-box terminology: "Confidence in climate model projections is based on physical understanding of the climate system and its representation in climate models…"

Within individual climate papers in the literature, a range of terminology is used for model types. Rahman and Lateh (2015) divide the range of methods into two broad groups: simulation techniques and statistical models. In their survey of methods, Adams et al. (2013) state that the two endpoints on a theoretical continuum of mechanisms are process-based and empirical model types. Moore et al. (2013) defined two model categories:  physically plausible models of reduced complexity that exploit statistical relationships between climate and climate forcing, and more complex physics-based models of the separate elements of the climate budget.

Again, none of the terms inductive, deductive, black-box or white-box were found.

As we are considering climate, we will use the terms *process-based* and *statistical* from the climate literature to stand for deductive/white-box and inductive/black-box respectively.

The key approach to check for model adequacy is termed (Montgomery et al. 2008) validation. A common method of validation is to check the ability of the simulation to correctly predict outputs caused by inputs other than those used in constructing the model. This is done by "saving" some system observations in order to use them later for validation. This is termed in statistics (Montgomery et al., 2008) the split-sample, cross-validation or train-test approach. Further discussion of validation is presented in Section 2 – Methods.

**1.2  The forecasting performance of existing process-based climate models**

Using the split-sample approach, Newman (2013) finds that an empirical time series model shows global surface temperature prediction skill better than that of the major process-based climate models (phase 5 of the Coupled Model Intercomparison Project (CMIP5)). To Newman, these results suggested that current coupled model decadal forecasts may not yet have much prediction skill beyond that captured by multivariate, predictably linear dynamics.

With regard to correct model specification, Grassi (2013) notes that process-based climate models (and too-simple empirical models) are often lacking in this regard due to the fact that climate series display complex statistical properties, and that modelling of these must be correctly specified to these to provide valid statistical inference.

In the IPCC AR5 chapter entitled Evaluation of Climate Models (IPCC 2013), Flato et al. state (Page 826) "…many studies have failed to find strong relationships between observables and projections"; and (Page 772): " Almost all CMIP5 historical simulations do not reproduce the observed recent warming hiatus."

One of the reasons for these problems when process models are used to model climate may be that, as Enting (2010) points out, process models are vulnerable to neglect of processes – the 'Kelvin error'. Enting writes: "The term 'Kelvin error' refers to the risk of missing a process from the modelling, taking its name from Lord Kelvin's underestimates of the ages of the earth and sun due to neglect of nuclear processes."

By contrast, a black-box model, while its component processes are not specified at all, by definition contains *all the component processes* of the reality under study.

Our previous study (Leggett and Ball 2015) utilised such a black-box approach, and explained the global surface temperature hiatus in terms of a Granger-causal relationship between first-difference atmospheric $CO_2$ and global surface temperature with a high level of statistical significance. This paper therefore will use the method of Leggett and Ball (2015) to forecast future global surface temperature.

### 1.3 Criteria for choosing the forecasting model for this paper

Within the realm of inductive (black-box) modelling, methods may be broken down initially into two main types – univariate and multivariate.

Multivariate modelling involves the utilisation of correlation between the outcome variable and at least one causal variable. Given the evidence for causality between atmospheric $CO_2$ and global surface temperature shown using multivariate modelling (in its bivariate form) (Leggett and Ball, 2015) the bivariate modelling approach is also used in this study.

Within the realm of bivariate modelling, the base model is prepared by regression analysis – often termed 'ordinary least squares analysis'.

Yet again, within this realm there are two broad sub-categories: ordinary least squares (OLS) regression, and regression corrected for a range of statistical issues which arise when the variables involved are time series. Time series models (Greene 2012) differ from ordinary regression models in that the sequence of measurements of a process over time introduces autocorrelations between measured values. The serial nature of the measurements must be addressed by careful examination of the lag structure of the model. When corrected, this type of OLS regression is termed 'time-series analysis'(Greene 2012).

Given the above, the choice between OLS and time-series analysis is straightforward when model specification is concerned – time series analysis is to be preferred. But when we turn to forecasting, these two aims can sometime be at odds. For example, regarding the choice between competing models Montgomery et al. (2008) write: "Concentrating on the model that produces the best historical fit often results in overfitting, or including too many parameters or terms in the model just because these additional terms improve the model fit. In general, the best approach is to select the model that results in the smallest standard deviation… of… forecast errors when the model is applied to data that was not used in the fitting process."

Stock (2001) also states that the most reliable way to evaluate a forecast or to compare forecasting methods is by examining out-of-sample performance.

Generally (Stock 2001; Montgomery et al. 2008), data that was not used in the fitting of the model is implemented by what is termed a split-sample or a cross-validation procedure. (Other terms for such methodologies are out-of-sample forecast, train-test methodology, and data splitting.) The term 'split-sample test' will be used in this paper.

Validation methodology is generally called 'detection' and 'attribution' in the climate change field (IPCC 2013).

### 1.4  Well-specified time series model

In this study, a number of bivariate time-series relationships are assessed. For each relationship, a well-specified bivariate time-series model requires that:

1. Each of the series used is stationary (at the least, trend-stationary (Greene 2012)). Concerning stationarity, a range of tests exists. In this study, to thoroughly seek stationarity, both the Augmented Dickey-Fuller (ADF) test and the Kwiatkowski-Phillips-Schmidt-Shin (KPSS) test are used. The KPSS test has more sensitivity and better discriminating power for series which are weakly stationary (Greene 2012).

2. A model is established in which any structural break in the relationship, if present, is fully accounted for (Clements and Hendry 1998). This is done by conducting in the first instance a CUSUM test as an initial indicator, and, if this indicates a break, using rolling Chow tests (Wooldridge 2009) to precisely show the timing of the break. It will be shown that this question is particularly relevant to global temperature series, and two representative, widely cited, temperature series are therefore assessed. One series is that of monthly instrumental temperature records for global surface temperature formed by combining the sea surface temperature records compiled by the Hadley Centre of the UK Met Office and the land surface air temperature records compiled by the Climatic Research Unit (CRU) of the University of East Anglia (Morice et al. 2012) (the HadCRUT4 version is used). The second series used is the UAH satellite temperature dataset, developed at the University of Alabama in Huntsville, infers the temperature of various atmospheric layers from satellite measurements of radiance (Christy et al. 2007). The lower troposphere series (termed here UAH) is used.

3. A model is established in which any autocorrelation in the relationship, if present, is fully accounted for. Autocorrelation is identified by conducting the Breusch–Godfrey LM test (Asteriou and Hall 2007), and, if present, is accounted for in the model by dynamic regression analysis (Greene 2012).

In summary for this section, we will seek a time series model which is both well specified and forecasts well. If, however, a less well specified model forecasts best in out-of-sample testing, we will use that model.

**1.5 Modeling future carbon dioxide sinks**

Two significant $CO_2$ sinks are present on earth which need to be taken into account in the modelling – the land (biota-based) sink and the ocean sink.

What constitutes reasonable modelling for the sinks? Some research has suggested that the contribution of sinks will decrease (Ciais et al. 2013), and some that it will increase (Schimel et al. 2014). Looking at the land sink, we note that plants have thrived previously – for example in the Carboniferous Period – under much higher global temperatures and $CO_2$ concentrations than at present or projected under any RCP scenario (Franck et al. 2006). In light of the foregoing, we conclude for the purposes of this paper that there is likely to be no threshold risk to the biosphere from any of the range of temperatures projected by the RCPs in the 21st century.

Similarly for the ocean chemical sink, Landschützer et al. (2015) report that while several studies had suggested that the carbon sink in the Southern Ocean – the ocean's strongest region for the uptake of anthropogenic $CO_2$ – has weakened in recent decades, the weakening trend stopped around 2002, and by 2012 the Southern Ocean had regained its expected strength, based on the growth of atmospheric $CO_2$. Hartin et al. (2015) also project the continued rise in the ocean carbon sink over the period of this paper's forecast, to 2100.

Given the above, our modelling assumes that the current relationships between each of the two sinks and atmospheric $CO_2$ will continue.

## 1.6 Forecasting global temperature to 2100

Two significant $CO_2$ sinks are present on earth which need to be taken into account in the modelling – the land (biota-based) sink and the ocean sink. In the paper, we use both time series assessment and split-sample testing for each stage of the modelling sequence in order to produce the best possible forecast.

We firstly forecast annual future performance of the CO2 sinks to 2100. Using data from Mohr et al (2015), we then forecast anthropogenic CO2 emissions under the peak fossil fuel scenario over the same period.

By deducting annual CO2 sink uptake from annual emissions, we can produce a forecast of net annual change in the level of atmospheric CO2. From this, we forecast future global surface temperature by year up to 2100.

The forecast future temperature trend from our modelling is then compared with the business-as-usual RCP temperature trend, RCP8.5 (Riahi et al. 2011).

## 2 Methods

### 2.1 Analysis sequence

The analysis sequence outlined in Sect. 1.6 above is operationalised as follows:

1. Obtain estimate of anthropogenic $CO_2$ emitted into the atmosphere per year (in gigatonnes of carbon) till 2100 under the peak fossil fuel scenario
2. Estimate land and ocean sinks uptake of $CO_2$ from the atmosphere per year to 2100 in gigatonnes of carbon
3. Estimate net $CO_2$ added to atmosphere per year in gigatonnes of carbon
4. Based on first-difference $CO_2$ and temperature relationship (Leggett and Ball 2015), forecast global surface temperature for each year to 2100.

In each case, the stationarity (see Section 1.4) of the time series is assessed and any autocorrelation in the model is accounted for. Next, the capacity of each model to forecast is assessed by the split sample method. For the temperature forecast, several sources of temperature data are used. For these models, forecasts are run for each temperature series and compared, with any convergence in results being noted.

Statistical methods used are standard (Greene 2012). Categories of methods used are normalisation; differentiation (approximated by differencing); and time-series analysis. To make it easier to assess the relationship between the key climate variables visually, data were normalised where required using statistical Z scores (also known as standard scores) (expressed as 'relative level' in the figures). In a Z-scored data series, each data point is part of an overall data series that sums to a zero mean and variance of 1, enabling comparison of data having different native units. Hence, when several Z-scored time series are depicted in a graph, all the time series will closely superimpose, enabling visual inspection to clearly discern the degree of similarity or dissimilarity between them. Individual figure legends contain details on the series lengths used to calculate the Z score.

All assessments were carried out using the time series statistical software package Gnu Regression, Econometrics and Time-series Library (GRETL) (available from: http://gretl.sourceforge.net/, accessed 30 June 2015).

Autocorrelation is identified by conducting the Breusch–Godfrey LM test (Asteriou and Hall 2007), and, if present, is accounted for in the model by dynamic regression analysis (Greene 2012) using the GRETL ARMAX procedure.

## 2.2 Data sources

For data sources for global temperature, we used the Hadley Centre–Climate Research Unit combined Landsat and SST surface temperature series (HadCRUT) version 4.2.0.0 (Morice et al., 2012), and for global lower troposphere temperature the UAH MSU/AMSU global satellite temperature dataset (UAH version 6.0) (Christy et al. 2007). For atmospheric $CO_2$, the US Department of Commerce National Oceanic and Atmospheric Administration Earth System Research Laboratory Global Monitoring Division Mauna Loa, Hawaii, annual $CO_2$ series (Keeling et al. 2009) is used. Data series relating to the Representative Concentration Pathway (RCP8.5) scenario model are from Riahi et al. (2011).

For the period 1959 to 2014, annual data on measured or estimated anthropogenic $CO_2$ emissions into the atmosphere, and $CO_2$ uptake from the atmosphere by land and ocean sinks in gigatonnes of carbon (GtC) are as estimated by the Global Carbon Project (Le Quéré et al. 2014).

Estimated fossil fuel emissions under the peak fossil fuel scenario are from the best-estimate scenario of Mohr et al. (2015) in gigatonnes of carbon per year (GtC/yr).

## 2.3 Presentation

It is noted that the forecasts do not include variance so actual annual outcomes will vary around the values forecast.

When points relating to the temperature slowdown are being made, Z scores are used from data start to 1998.

To minimise visual clutter in figures displaying multiple curves, error bars are not included. To illustrate the error performance of the key findings, error bars are shown in the concluding figures.

In the tables of results, statistical significance at the 10%, 5% and 1% levels is indicated by *, **, and *** respectively.

## 3 Results

### 3.1 Forecast of land and ocean sinks to 2100

In this section we estimate the atmospheric $CO_2$ series that would be generated by the peak fossil fuel emissions scenario. As discussed, the net $CO_2$ in the atmosphere results from emissions into the atmosphere, some of which are absorbed by land and ocean sinks, with the remaining amount accumulating by year. Figure 1 shows trends for each of these variables in gigatonnes of carbon per year from 1959 to 2014. The measured or estimated carbon uptake per year for these variables is as estimated by the Global Carbon Project (Le Quéré et al. 2014).

It can be seen that there are two main anthropogenic sources of $CO_2$ emissions – (i) fossil fuel and cement; and (ii) land use change. There are two main sinks for $CO_2$ emissions – land and ocean.

#### 3.1.1 Future emissions and sink performance

Fossil fuel $CO_2$ emissions are from the best-estimate peak fossil fuel scenario of Mohr et al. (2015) in GtC/yr. Current cement and land use change emissions are small in comparison with those from fossil fuel (Le Quéré et al. 2014); for the purposes of this study they are considered neither to increase on nor decrease from current levels. Figure 2 shows the resulting trend to 2100 in gigatonnes per year for the anthropogenic emissions.

In this study, sink uptake is modelled as the sum of the uptake of the land and ocean sinks. Figure 3 shows the total $CO_2$ uptake of the land and ocean sinks from 1959 to 2013 in GtC/yr.

We now turn to estimating the $CO_2$ uptake performance of the land and ocean sinks to 2100.

Figure 4 shows the relative scale and relative trend of the anthropogenic emissions and the combined sinks (Le Quéré et al. 2014). It can be seen that sink uptake has risen as anthropogenic emissions have risen.

The forecast of sink performance is sought in the first instance using time series analysis and, as discussed in Section 1.4, two requirements must be met. These are: (i) that series are stationary; and (ii) that any autocorrelation in the relationship between the two series is identified, specified, and fully taken into account.

### 3.1.2. Assessing stationarity of sink and anthropogenic $CO_2$ series

As stated above, all series used in a time-series regression must be stationary (Greene 2012). Here it should be noted that trend stationarity is adequate stationarity for the present form of analysis (Greene 2012). To assess the stationarity of the following series we use both the augmented Dickey-Fuller (ADF) test and the Kwiatkowski–Phillips–Schmidt–Shin (KPSS) test. The KPSS test and the ADF test complement each other to provide a fuller range of sensitivity to stationarity than either would if used on its own (Kwiatkowski et al. 1992). This is especially important as it is known that the ADF test has relatively low power in many situations.

Hence, the stationarity test protocol for this paper is that non-stationarity is demonstrated only if results of both ADF and KPSS tests show lack of stationarity.

The results of the tests are given in Table 1.

The above testing shows that neither of the two series is shown by both ADF and KPSS tests to exhibit lack of trend stationarity. Hence each series is considered trend stationary and suitable for use in time-series analysis (Greene 2012).

Therefore we next assess the quality of the forecast from the regression between the anthropogenic $CO_2$ series and the sink series. This assessment is done using the split-sample test process.

### 3.1.3 Split-sample test for forecast of sink from anthropogenic $CO_2$ emissions

In this section the quality of the time series model for training period of split-sample test is assessed.

Table 2 shows that, for the training period selected for the split-sample test – 1959 to 1999 – a basic OLS model is well specified: there is a correlation between the two variables which is statistically significant, there is no autocorrelation (LM test), and there is no evidence (CUSUM test) of a structural break over the period.

Figure 5 depicts the forecast result in comparison with observed data for the same period.

Two assessments of the results of the above split-sample test are now made: one based on correlations, and one based on Theil's U2 statistic.

Table 3 shows the extent of correlation between (i) the sink performance for the test period forecast from the training period relationship, and (ii) the observed sink performance for the test period.

Table 3 shows that the fit of the regression is highly statistically significant, and that there is no evidence of autocorrelation.

Theil's inequality coefficient in its U2 form provides a measure of the distance of the true values from the forecast values and is a widely cited single measure of forecast accuracy (Watson and Teelucksingh 2002). The Theil's U2 value for the above relationship is 0.70.

The U2 statistic will take the value 1 under the naïve forecasting method (one predicting no change) (Watson and Teelucksingh 2002). Values less than 1 indicate greater forecasting accuracy than the naïve forecasts, values greater than 1 indicate the opposite. This means that the above model predicting sink uptake from anthropogenic $CO_2$ emissions forecasts better than naïve.

Further, Table B3 of ACT Department of Treasury (2008) provides information on the performance of mainstream official Australian Treasury models. Inspection of the table shows that the above split-sample test prediction of sink uptake from anthropogenic $CO_2$ emissions performs comparably to such models.

**3.1.4 Forecast of sink performance to 2100**

Given the above combination of statistically significant forecasting in the split-sample test setting and the Theil's U2 result, we will now use the model to forecast sink carbon uptake performance to 2100.

Table 4 shows a straightforward OLS model from 1959 to 2014 for the series which will be used to create the sink forecast to 2100.

The table shows that autocorrelation is adequately accounted for in the straightforward OLS model (LM test *p*-value = 0.83) so that it can be used in forecasting without the need for further processing to account for autocorrelation.

As stated in Section 3.1.1, the forecast uses the Mohr et al. (2015) best-estimate peak fossil fuel scenario as the data for the independent variable from 2013 to 2100 for the forecast of sink performance. The results of the forecast for sink performance from this model are given in Figure 6.

Figure 6, then, shows the expected removal of $CO_2$ per year (in gigatonnes) from the atmosphere by the land and ocean sinks. In the next section this is deducted from forecast anthropogenic emissions into the atmosphere to give us the net amount of $CO_2$ added to the atmosphere per year.

**3.2 Estimate of net $CO_2$ added to atmosphere per year**

The $CO_2$ per year added to the atmosphere is derived by subtracting the amount of $CO_2$ absorbed by the sinks shown in Figure 6 from the amount of $CO_2$ emitted by anthropogenic sources. These series and the result are illustrated in Figure 7.

The time series of first-difference cumulative net $CO_2$ remaining in the atmosphere is required for use in a forecast based on the Leggett and Ball (2015) first-difference model. This time series is already available – the first-difference of cumulative net $CO_2$ is equivalent to the *level* of *annual* net $CO_2$ (green curve in Figure 7).

For what follows, observed first-difference annual atmospheric $CO_2$ to 2015 is spliced with forecast annual net $CO_2$ (equivalent to first-difference cumulated net $CO_2$) added to the atmosphere annually from 2015 to 2100.

**3.3  Estimate of global surface temperature for each year**

In the following sections we will use the above series to forecast future average global temperature from the atmospheric $CO_2$ scenario depicted in Figure 7. First, the stationarity or otherwise of the series used must be assessed.

**3.3.1 Assessing stationarity of the series**

Table 5 displays the results of ADF and KPSS tests for trend stationarity for the above series allowing for both drift and trend.

Table 5 shows that none of the above series can be shown by both ADF and KPSS tests to exhibit lack of trend stationarity. Hence each series is considered trend stationary and suitable for use in time-series analysis.

We now assess the relationship between the first difference $CO_2$ series and each global temperature series. The first step in this assessment is testing the relationship for the presence of structural breaks.

**3.3.2  Relationship between first-difference atmospheric $CO_2$ and temperature: test for structural break**

To test for any structural breaks in the relationship between first-difference atmospheric $CO_2$ and temperature (UAH and HadCRUT4), rolling Chow tests are conducted for each of the two relationships. The results are shown for each of the two temperature series in Figure 8.

Figure 8 shows that the largest Chow test result for the relationship of first-difference $CO_2$ with each of UAH and Hadcrut4 is in the same year, 1989. From here on, modelling is carried out to account for this structural break by means of a dummy variable from the start of the series (1960 for Hadcrut4, and 1979 for UAH) to 1989.

In the next section, the forecast performance of each of the two models – of first-difference $CO_2$ with UAH and HadCRUT4 respectively – is assessed using the split-sample test protocol (Montgomery et al. 2008).

For the split-sample assessment, data series pairs to 2015 are divided into a training sample from the start of data to 1999, and a test sample from 2000 to 2015.

Table 6 presents a straightforward OLS model for first-difference $CO_2$ and HadCRUT4 for the training phase of the model using data from 1959 to 1999 (including a dummy variable for the structural break described in Sect. 3.3.2 above).

The table shows that the model to 1999 is quite well specified. There is no evidence of a further structural break (CUSUM test); and there is only a small amount of autocorrelation (LM test).

This autocorrelation is addressed in Table 7 by means of a dynamic regression model. The Ljung-Box test statistic reported in Table 7 shows that the autocorrelation issue is now addressed.

We now turn to the case for first-difference $CO_2$ and UAH.

In Table 8, the first-difference $CO_2$-UAH relationship over the split-sample test training period of 1979 to 1999 is assessed for adequacy of time-series specification.

It can be seen from Table 8 that the first-difference $CO_2$-UAH relationship is well specified using the straightforward OLS model.

All three models are now assessed in the testing phase of the split-sample test procedure.

Figure 9 shows the forecasting performance of the two first-difference $CO_2$–HadCRUT4 models developed above in the training phase of the split-sample test procedure.

In Figure 9, the red curve depicts the output of the OLS model shown in Table 6; the green curve is for the ARMAX dynamic regression model shown in Table 7.
It can be seen that both models agree closely with each other. However each model underestimates the reported HadCRUT4 time series for the test period.

Figure 10 shows the forecasting performance of the first-difference $CO_2$–UAH model developed above in the training phase of the split-sample test procedure.

Comparison of Figures 9 and 10 shows that the UAH training relationship performs much better in forecasting the reported temperature in the test period 2000-2015 than the HadCRUT4 relationship. The HadCRUT4 reported series for the test period shows the same signature as the test series but is elevated above it. There is then a further elevation for reported 2014 and 2015 HadCRUT4 data.

Theil's U2 data concerning the differences in forecasting performance are captured in Table 9. As mentioned above, the U2 statistic will take the value 1 under the naïve forecasting method. Values less than 1 indicate greater forecasting accuracy than the naïve forecasts, values greater than 1 indicate the opposite (Watson and Teelucksingh 2002). This means that the UAH model forecast better than naïve, the HadCRUT4 worse. Table B3 in Australian Treasury (2008) provides information on the performance of mainstream Australian Treasury models. It can be seen that the UAH forecast performs very comparably to those models.

In the light of the foregoing, we firstly seek convergence in the forecasts from models using each of UAH and HadCRUT4. If convergence is not found, the UAH model will be taken as likely to be the more accurate.

**3.4 Forecast temperature to 2100**

Having established in Section 3.2 the peak fossil fuel trajectory to be used in the forecast of temperature to 2100, we now turn to the base-case comparison to be used This is the scenario in which no peak in fossil fuel occurs. The rationale for selecting the base case is as follows.

The Mohr et al. (2015) peak fossil fuel scenario is of fossil fuel availability in a setting in which no particular policies are brought in to address the trend. Supply and demand are left simply to operate – that is, business-as-usual. Of the four IPCC RCP scenarios, only one is under such business-as-usual conditions: RCP8.5. From Riahi et al. (2011): "Compared to the scenario literature RCP8.5 depicts thus a relatively conservative business as usual case with low income, high population and high energy demand due to only modest improvements in energy intensity."

Riahi et al. also note that in RCP8.5 the extra demand is primarily met by fossil fuels.

Figure 11 depicts RCP8.5 emissions to the atmosphere in gigatonnes per year (Riahi et al., 2011) in comparison to the peak fossil fuel emissions trajectory (Mohr et al., 2015) previously illustrated in Figure 4. It can be seen that the peak fossil fuel estimate is much lower.

The temperature arising from the RCP8.5 emissions trajectory is available from the literature (Riahi et al. 2011). This temperature trend is therefore used directly to compare with the forecast temperature for the first-difference peak fossil fuel case made in the next sections.

A number of combinations make up the results. These are: first-difference RCP8.5 without fossil fuel peaking (this temperature trend is taken from the literature (Riahi et al. 2011)); peak fossil fuel if linear (this model is not run due to lack of stationarity of the cumulated peak fossil fuel series); first-difference $CO_2$ from the RCP8.5 atmospheric $CO_2$ estimate (taken from the literature); and first-difference $CO_2$ from the peak fossil fuel emissions scenario. We will deal with the two categories for which models were run in turn.

### 3.4.1 Forecast based on first-difference RCP8.5

This section provides the results for forecasts based on the relationship between first-difference RCP8.5 atmospheric $CO_2$ to 2100 (that is, without fossil fuel peaking) and temperature using each of the HadCRUT4 and UAH temperature series. Tables 10, 11, and 12 present the results of the modelling on which the forecasting is based; the forecasts are given in Figure 12.

The temperature forecasts arising from using these models are shown in Figure 12.

The figure shows that, by 2030, both HadCRUT4 models are forecasting closely similar results. This is in some contrast to the model for UAH, which, it is recalled, was the best model in terms of specification and split-sample test performance, and which forecasts a higher temperature trajectory. Overall, however, the figure shows that all first-difference RCP8.5 three models, as expected, forecast a lower temperature than expected from the RCP8.5 temperature trajectory based on level of $CO_2$.

### 3.4.2 Forecast based on first-difference $CO_2$ trajectory to 2100 derived from peak fossil fuel scenario

Three models are run. The first is an OLS model using UAH. This displays no autocorrelation or parameter instability (Table 13 ) and so is selected for forecasting. The second model also utilises OLS and involves HadCRUT4. This initial HadCRUT4 model displays autocorrelation and parameter instability (Table 14). A further, ARMAX, model for HadCRUT4 is therefore prepared. This model, of 2,0,0 form, displays no autocorrelation (Table 15).

Forecasts using the three models are given in Figure 13. In the figure, and those which follow, series are Z-scored from 1979 to 1998. The HadCRUT4 series is shown in absolute terms for comparison and scaling purposes.

In Figure 13 the vertical scale is adjusted in comparison with Figure 12 to enable concentration on the details of forecasts based on the first-difference $CO_2$ relationships. The figure shows that, as expected, all three temperature forecasts are markedly different from the RCP8.5 forecast. Further, all three forecasts tend to be similar, especially compared with that for RCP8.5.

The key finding is that temperature is forecast to continue to rise, gently, until around 2023 to slightly higher values than at present and then display an also gentle steady decrease.

In Figures 14 and 15 forecasts for UAH and HadCRUT4 respectively are shown with error bars and compared with the RCP8.5 projection.

As noted, the RCP8.5 temperature projection is an output of climate Global Circulation Models (GCMs). As discussed, climate GCMs and Earth System Models are examples of process-based models, which are deterministic models and so provide only point estimates in any individual forecast (Enting 2010; Stephenson et al. 2012). For this reason, Stephenson et al. write: "Because climate models do not themselves produce probabilities, an additional level of explicit probabilistic modelling is necessary to achieve this. … However, there is little agreement on what is the most reliable and robust methodology for making probabilistic predictions of real-world climate." A probabilistic model is needed to produce confidence limits. Standard-deviation-like depictions occur in graphs of temperature projections in the most recent IPCC assessment report (IPCC 2013) – for example, Figure 12.5 (page 1054). However in text on the same page, it is noted that these depict averages of multiple model runs, not estimates of the statistical uncertainty present in a single statistical model.

For the above reasons, the RCP data in Figures 14 and 15 are presented without confidence limits. Nonetheless the figure shows that even without error bars on the RCP8.5 projection, the UAH forecast and the RCP8.5 projection are along such substantially different trajectories that they are highly likely to be statistically significantly different.

## 4 Discussion

The aim of this paper was to forecast a global surface temperature trajectory to 2100 resulting from the effect of two future scenarios that were not explicitly in the range of scenarios (the Representative Concentration Pathway scenarios) utilised by the IPCC. These two scenarios are the emissions trend under peak fossil fuel (for example, Mohr et al., 2015); and the climate sensitivity determinable from the relationship between first-difference $CO_2$ and temperature recently shown by Leggett and Ball (2015).

In carrying out this work, we have attempted to provide the justification for our choice of statistical model to use by means of a comprehensive scan of the range of models available and of their strengths and weaknesses.

Similarly we have attempted to lay out a comprehensive approach both to within-model performance and to the trade-off between a well-specified time-series model and a model which forecasts well.

With this background we found that in general, the models we used - either well specified or less so - converged in their forecasts, so adding to the potential robustness of the findings.

All models utilised displayed high statistical significance.

Under the effect of the combination of the peak fossil fuel and first-difference $CO_2$ scenarios, we found that temperature is forecast to continue to rise, but only gently, until around 2023 where it reaches a level slightly higher than at present; and from then to display an also gentle steady decrease. It is shown that this trajectory is markedly lower than that generated by the IPCC business-as-usual level-of-$CO_2$ (RCP8.5) model (Riahi et al. 2011).

These lower results are evidence that the climate problem may require less future preventative action than is presently being considered necessary. If so, it is stressed that the same evidence is support for the case that the peak fossil fuel problem (Mohr et al. 2015) would require ameliorative action. This globally required action is the same as it would have been for climate* – the rapid transition to a predominantly renewable global energy system.

-------------------------------------------------------------------------------------------

*The draft Paris Agreement (United Nations 2015) is the draft global agreement achieved within the framework of the United Nations Framework Convention on Climate Change (UNFCCC) with the goal of reducing global warming and aimed to come into force in the year 2020. The agreement was negotiated by representatives of 195 countries at the 21st Conference of the Parties of the UNFCCC in Paris and adopted by consensus on 12 December 2015. The draft agreement recognises at the outset (page 1) that "deep reductions in global emissions will be required"; the UNFCCC states separately (United Nations 2016) that "an emphasis on renewable energy sources will be essential for achieving this goal."

Table 1: Results of tests for trend stationarity allowing for both drift and trend, for level of anthropogenic $CO_2$ and for land plus ocean sinks. Data 1959-2013; T=56

|  | ADF test statistic | *p*-value | ADF test interpretation | KPSS test statistic (Lag truncation parameter = 3) | *p*-value | KPSS test interpretation | Overall stationarity assessment |
|---|---|---|---|---|---|---|---|
| Anthropogenic $CO_2$ emissions | -1.91337 | 0.6341 | Non-stationary | 0.120227 | > .10 | Stationary | Stationary |
| Land plus ocean sinks $CO_2$ uptake | -6.3951 | 8.47E-06 | Stationary | 0.0576965 | > .10 | Stationary | Stationary |

Table 2: OLS, using observations 1959-1999 (T = 41). Dependent variable: Land and ocean sink $CO_2$ uptake

|  | Coefficient | *p*-value | Statistical significance |
|---|---|---|---|
| const | 0.468732 | 0.546 |  |
| Anthropogenic $CO_2$ emissions | 0.492828 | 0.0002 | *** |

|  | *p*-value |
|---|---|
| LM test for autocorrelation up to order 3 (Null hypothesis: no autocorrelation) | 0.863724 |

|  | Test statistic: Harvey-Collier t(17) *p*-value |
|---|---|
|  |  |

| CUSUM test for parameter stability Null hypothesis: no change in parameters | 0.899626 |

Table 3. OLS, using observations 2000-2014 (T = 15). Dependent variable: Observed Sinks $CO_2$ uptake (GtC/yr)

|  | Coefficient | *p*-value | Statistical significance |
|---|---|---|---|
| const | −2.91474 | 0.2179 |  |
| Sink $CO_2$ uptake (GtC/yr) forecast from anthropogenic $CO_2$ emissions | 1.58539 | 0.0033 | *** |

|  | *p*-value |
|---|---|
| LM test for autocorrelation up to order 3 (Null hypothesis: no autocorrelation) | 0.673783 |

Table 4: OLS, using observations 1959-2014 (T = 56). Dependent variable: Sink $CO_2$ uptake (GtC/yr)

|  | Coefficient | p-value | Statistical significance |
|---|---|---|---|
| const | 0.252277 | 0.6249 |  |
| Observed anthropogenic $CO_2$ emissions (GtC/yr) | 0.525473 | 5.56E-10 | *** |

|  | p-value |
|---|---|
| LM test for autocorrelation up to order 3 (Null hypothesis: no autocorrelation) | 0.834669 |

Table 5: Results of tests for trend stationarity allowing for both drift and trend, for atmospheric $CO_2$ forecast to 2100, global surface temperature (HadCRUT4) and global lower tropospheric temperature (UAH)

|  | Data details | ADF test statistic | $p$-value | ADF test inter-pretation | KPSS test statistic (Lag truncation parameter = 3) | $p$-value | KPSS test inter-pretation | Overall stationarity assessment |
|---|---|---|---|---|---|---|---|---|
| Atmospheric $CO_2$ forecast to 2100 | Data 1959-2013; T=56 | -5.378 | 0.0005 | Stationary | 0.0588 | >.10 | Stationary | Stationary |
| Global surface temperature (HadCRUT4) |  | -4.656 | 0.027 | Stationary | 0.2065 | 0.014 | Non-stationary | Stationary |
| Global lower tropospheric temperature (UAH) | Data 1979-2013; T = 34 | -5.668 | 0.0002 | Stationary | 0.0959 | > .10 | Stationary | Stationary |

Table 6: OLS, using observations 1960-1999 (T = 40). Dependent variable: HadCRUT4 (degrees Celsius)

|  | Coefficient | *p*-value | Statistical significance |
|---|---|---|---|
| const | −0.000899668 | 0.9951 |  |
| First-difference $CO_2$ | 0.403961 | <0.0001 | *** |
| Dummy variable to 1989 | −1.22928 | <0.0001 | *** |

|  | Test statistic: Chi-square: *p*-value |
|---|---|
| LM test for autocorrelation up to order 3 (Null hypothesis: no autocorrelation) | 0.0452719 |

Table 7. ARMAX dynamic regression, using observations 1960-1999 (T = 40). Dependent variable: Z HadCRUT4

|  | Coefficient | *p*-value | Statistical significance |
|---|---|---|---|
| const | 0.000638168 | 0.9975 |  |
| HadCRUT led 1 year | 0.0904103 | 0.5404 |  |
| HadCRUT led 2 years | 0.368453 | 0.0131 | ** |
| First-difference $CO_2$ | 0.406185 | 3.74e-011 | *** |
| Dummy variable to 1989 | −1.20733 | 5.24e-08 | *** |

|  | Test statistic: Chi-square: *p*-value |
|---|---|
| Ljung-Box Q' test for autocorrelation up to order 3 (Null hypothesis: no autocorrelation) | 0.2114 |

Table 8: OLS, using observations 1979-1999 (T = 21). Dependent variable: Z Global lower tropospheric temperature (UAH)

|  | Coefficient | p-value | Statistical significance |
|---|---|---|---|
| const | 0.110217 | 0.459 |  |
| Z First-difference $CO_2$ | 0.711487 | <0.0001 | *** |
| Dummy variable to 1989 | −0.806896 | 0.0005 | *** |

|  | p-value |
|---|---|
| LM test for autocorrelation up to order 3 (Null hypothesis: no autocorrelation) | 0.58412 |

|  | Test statistic: Harvey-Collier t(17) p-value |
|---|---|
| CUSUM test for parameter stability Null hypothesis: no change in parameters | 0.471736 |

Table 9. Theil's U2 Forecast evaluation statistics for modelled time-series

| | ZHadCRUT4, OLS modelled to 1999 from first-difference $CO_2$ with dummy variable to 1989 (slight autocorrelation remaining) | ZHadCRUT4, ARMAX 2,0,0 modelled to 1999 from first-difference $CO_2$ with dummy variable to 1989 (no autocorrelation remaining) | Z UAH, OLS modelled to 1999 from first-difference $CO_2$ with dummy variable to 1989 (no autocorrelation remaining) |
|---|---|---|---|
| Theil's U2 | 2.19 | 2.14 | 0.55 |

Table 10: OLS, using observations 1960-2015 (T = 56). Dependent variable: HadCRUT4

|  | Coefficient | p-value | Statistical significance |
|---|---|---|---|
| const | 14.3475 | 1.19E-107 | *** |
| Z first-difference $CO_2$ to 2100 (RCP8.5 scenario) | 0.111724 | 2.89E-10 | *** |
| Dummy variable to 1989 | −0.308075 | 6.45E-15 | *** |

|  | p-value |
|---|---|
| LM test for autocorrelation up to order 3 (Null hypothesis: no autocorrelation) | 0.0189164 |

|  | Test statistic: Harvey- |
|---|---|
|  |  |

|  | Collier t(17) p-value |
|---|---|
|  |  |

Table 11: ARMAX dynamic regression, using observations 1960-2015 (T = 56) Dependent variable: HadCRUT4 (degrees Celsius)

|  | Coefficient | p-value | Statistical significance |
|---|---|---|---|
| const | 14.3558 | 0 | *** |
| HadCRUT led 1 year | 0.246884 | 0.0814 | * |
| HadCRUT led 2 years | 0.370035 | 0.0077 | *** |
| Z first-difference $CO_2$ to 2100 (RCP8.5 scenario) | 0.0944603 | 5.64E-12 | *** |
| Dummy variable to1989 | −0.304469 | 7.99E-10 | *** |

|  | Test statistic: Chi-square |
|---|---|
|  |  |
| LM test for autocorrelation up to order 3 (Null hypothesis: no autocorrelation) | 0.973269 |

Table 12: OLS, using observations 1979-2015 (T = 37). Dependent variable: UAH

| | Coefficient | *p*-value | **Statistical significance** |
|---|---|---|---|
| const | 1.3201 | 9.09E-11 | *** |
| Z first-difference $CO_2$ to 2100 (RCP8.5 scenario) | 1.25786 | 3.54E-11 | *** |
| Dummy variable to 1989 | −1.50714 | 2.50E-07 | *** |

| | *p*-value |
|---|---|
| LM test for autocorrelation up to order 3 (Null hypothesis: no autocorrelation) | 0.760595 |

|  | Test statistic: Harvey-Collier t(17) p-value |
|---|---|
| CUSUM test for parameter stability Null hypothesis: no change in parameters | 0.705168 |

Table 13: OLS, using observations 1979-2015 (T = 37). Dependent variable: UAH

|  | Coefficient | p-value | Statistical significance |
|---|---|---|---|
| const | 1.3201 | <0.0001 | *** |
| First-difference $CO_2$ | 1.25786 | <0.0001 | *** |
| Dummy variable to 1989 | −1.50714 | <0.0001 | *** |

|  | p-value |
|---|---|
| LM test for autocorrelation up to order 3 (Null hypothesis: no autocorrelation) | 0.390143 |

|  | Test statistic: Harvey-Collier t(17) *p*-value |
|---|---|
| CUSUM test for parameter stability Null hypothesis: no change in parameters | 0.760595 |

Table 14: OLS, using observations 1960-2015 (T = 56). Dependent variable: HadCRUT4

|  | Coefficient | *p*-value | Statistical significance |
|---|---|---|---|
| const | 14.3475 | <0.0001 | *** |
| First-difference $CO_2$ | 0.111724 | <0.0001 | *** |
| Dummy variable to 1989 | −0.308075 | <0.0001 | *** |

|  | *p*-value |
|---|---|
| LM test for autocorrelation up to order 3 (Null hypothesis: no autocorrelation) | 0.0189164 |

|  | Test statistic: Harvey-Collier t(17) p-value |
|---|---|
| CUSUM test for parameter stability Null hypothesis: no change in parameters | 0.0763131 |

Table 15: ARMAX dynamic regression, using observations 1960-2015 (T = 56). Dependent variable: HadCRUT4

|  | Coefficient | p-value | Statistical significance |
|---|---|---|---|
| const | 14.3558 | <0.0001 | *** |
| HadCRUT4 led 1 yr | 0.246884 | 0.0814 | * |
| HadCRUT4 led 2 yrs | 0.370035 | 0.0077 | *** |
| First-difference $CO_2$ | 0.0944603 | <0.0001 | *** |
| Dummy variable to1989 | −0.304469 | <0.0001 | *** |

|  | p-value |
|---|---|
| LM test for autocorrelation up to order 3 (Null hypothesis: no autocorrelation) | 0.973269 |

| | Test statistic: Harvey-Collier t(17) p-value |
|---|---|
| CUSUM test for parameter stability Null hypothesis: no change in parameters | 0.3239 |

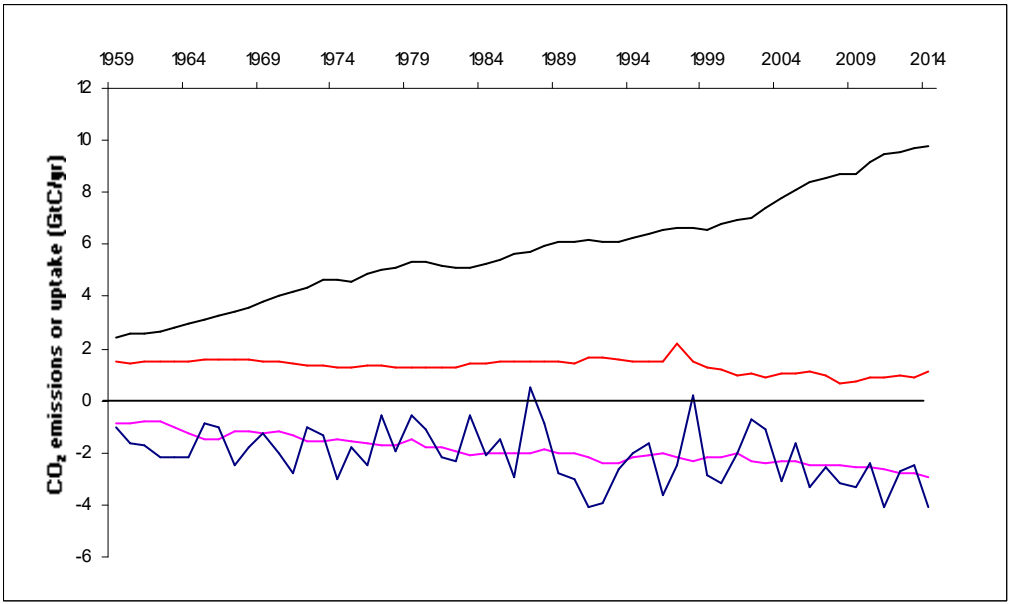

**Figure 1. Carbon dioxide emitted to or taken up from the atmosphere in gigatonnes of carbon per year (GtC/yr) from: fossil fuel and cement (black curve); land use change (red curve); land sink (sign reversed) (blue curve); ocean sink (sign reversed) (purple curve)**

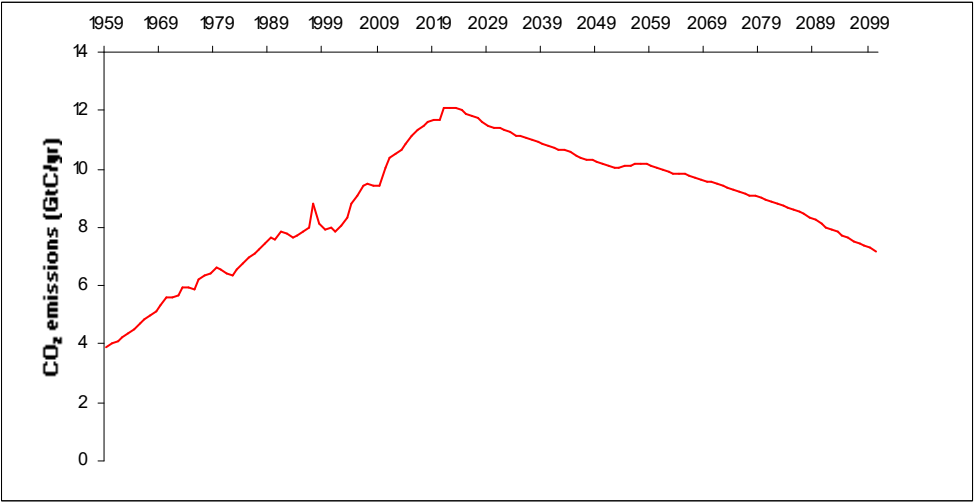

**Figure 2: Anthropogenic $CO_2$ emissions (from fossil fuel, cement and land-use change) (GtC/yr): from 1959-2014: observed (Le Quéré et al., 2014); from 2015-2100: best estimate from Mohr et al. (2015)**

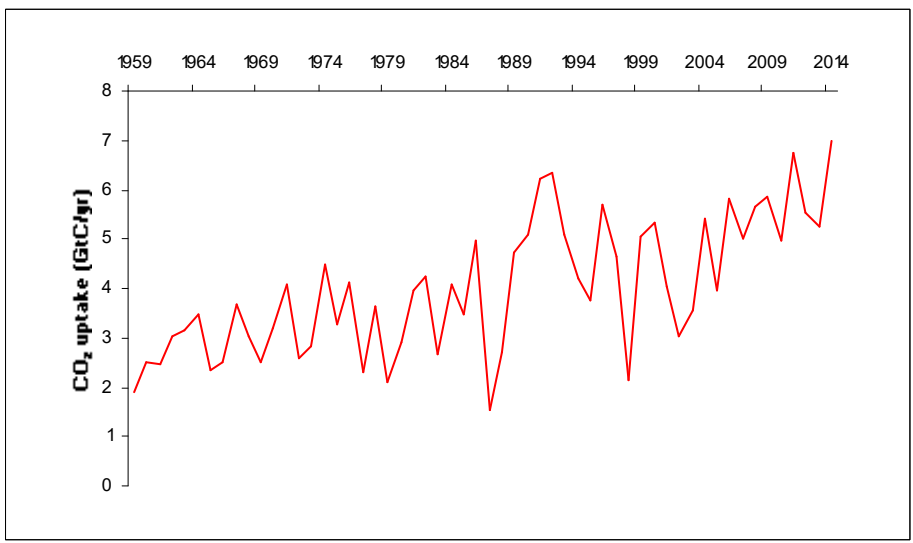

**Figure 3. Total CO$_2$ uptake of land sink and ocean sink (GtC/yr)**

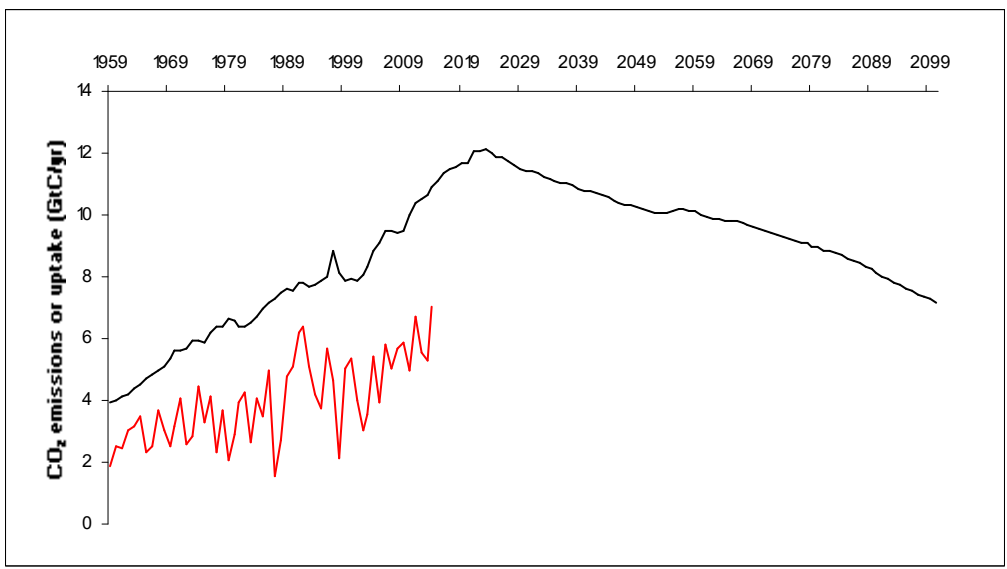

**Figure 4. Comparison of anthropogenic CO$_2$ emissions to 2100 under peak fossil fuel scenario (Mohr et al., 2015) (black curve) and total CO$_2$ uptake of land and ocean sinks (red curve) in GtC/yr**

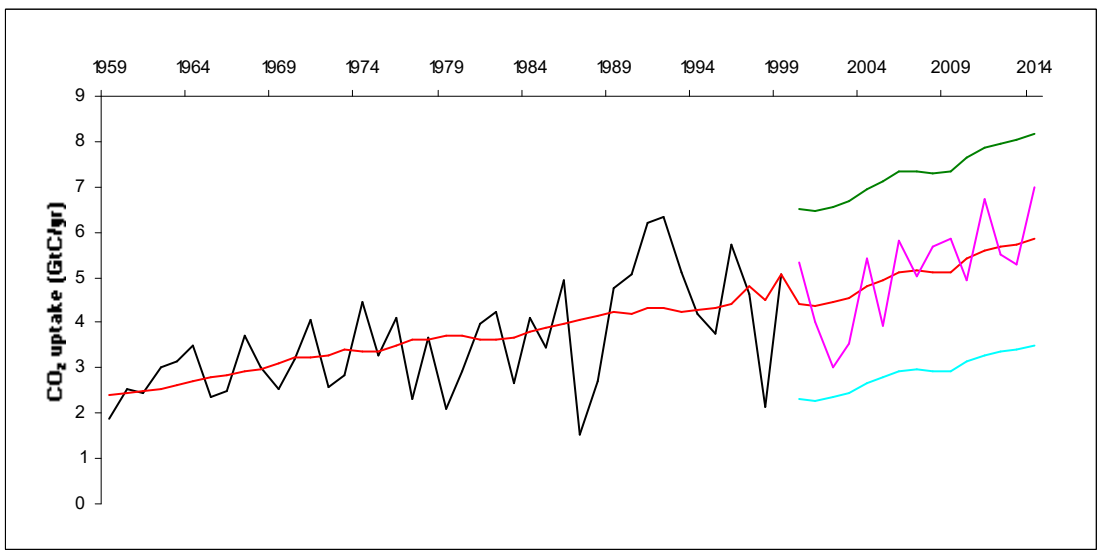

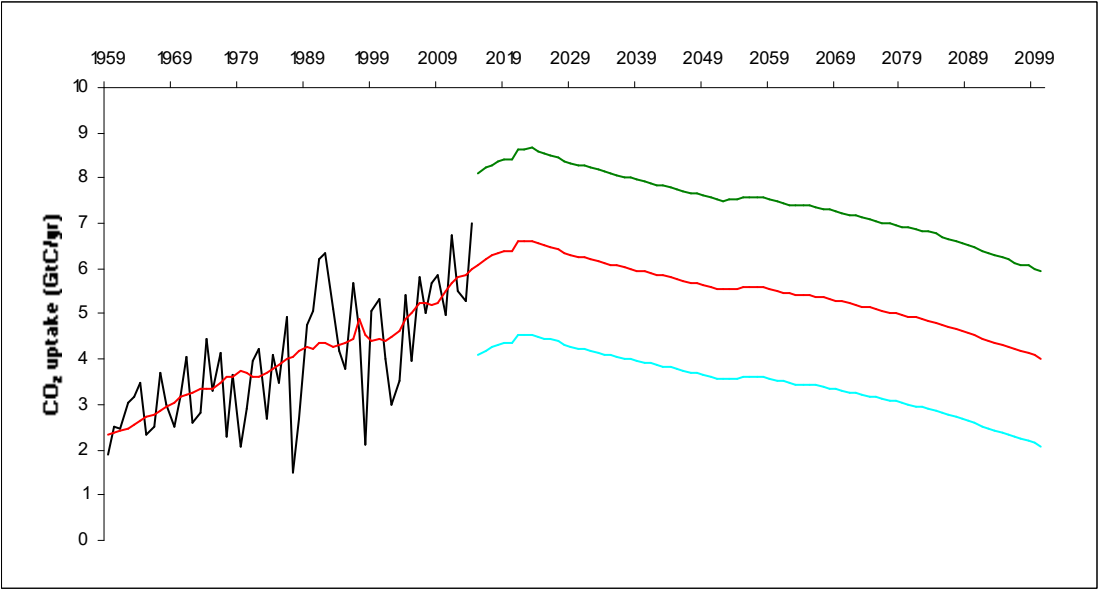

**Figure 5.** Land and ocean sink $CO_2$ uptake (GtC/yr): 1959-1999 (black curve); 2000-2014 (purple curve) (Le Quéré et al., 2014); forecast from observed anthropogenic $CO_2$ emissions 1959-1999 (red curve); 95% confidence limits for forecast (green and turquoise curves).

**Figure 6.** Land and ocean sink $CO_2$ uptake: 1959-2014 (black curve) (Le Quéré et al., 2014); forecast from observed anthropogenic $CO_2$ emissions (Le Quéré et al., 2014) to 2014; then forecast from 2015 to 2100 from anthropogenic $CO_2$ emissions 1959-2100 under peak fossil fuel scenario (Mohr et al., 2015) (red curve). 95% confidence limits for forecast: green and turquoise curves.

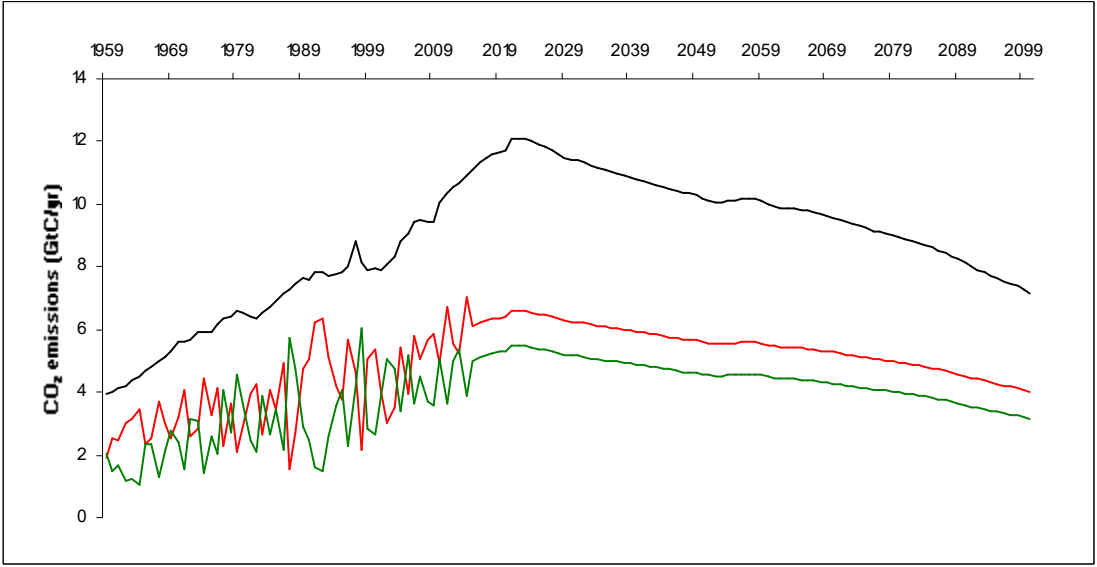

**Figure 7.** Anthropogenic $CO_2$ emissions forecast to 2100 under peak fossil fuel scenario (a) (black curve); land and ocean sink $CO_2$ uptake: observed then forecast 1959-2100 (b) (red curve); nett additional atmospheric $CO_2$ (a minus b) (green curve). All GtC/yr

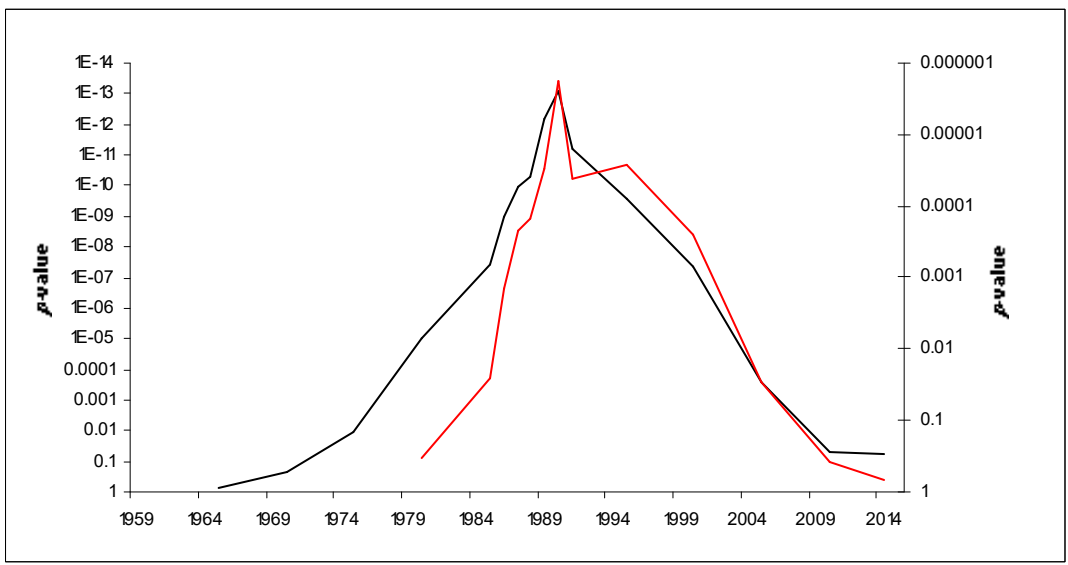

**Figure 8.** Results (reverse *p*-values) for rolling Chow tests for relationships between first-difference atmospheric $CO_2$ and global temperature – HadCRUT4 (black curve) and UAH (red curve )

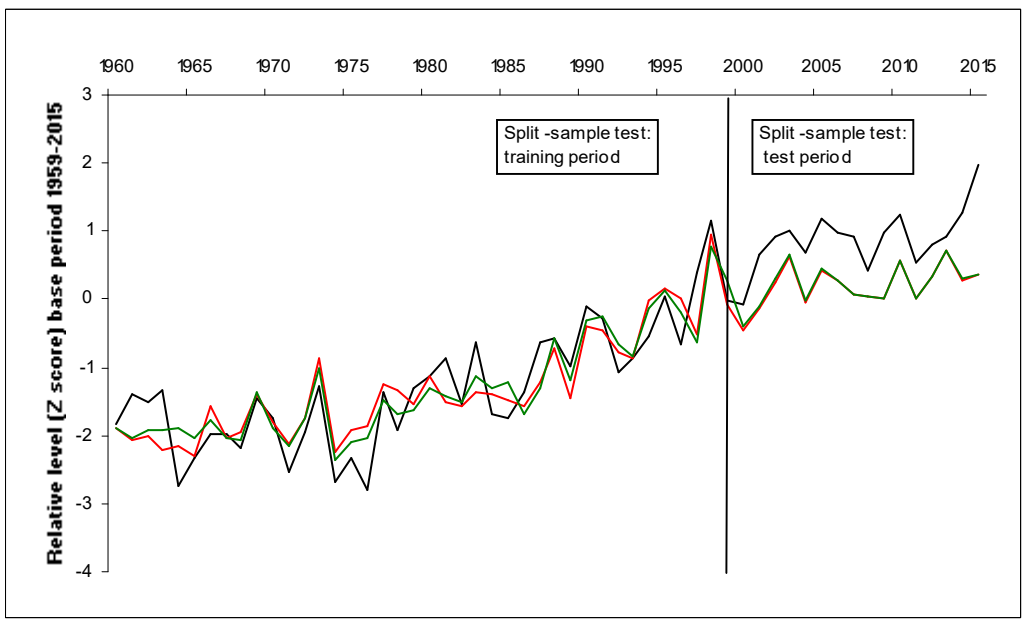

**Figure 9: Z scores. Split sample test: results for relationship between first-difference $CO_2$ and HadCRUT4. Training period 1960-1999; test period 2000-2015. Reported HadCRUT4 data: black curve. Results: OLS model shown in Table 6 (red curve); dynamic regression model shown in Table 7 (green curve).**

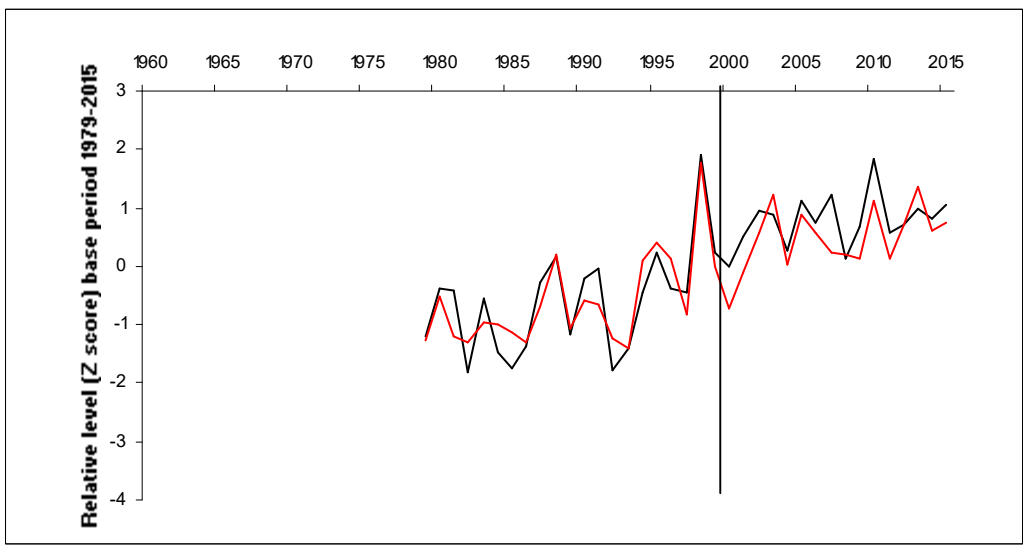

**Figure 10: Z scores. Split sample test: results for relationship between first-difference $CO_2$ and UAH. Training period 1979-1999; test period 2000-2015. Reported UAH data: black curve; train-test model result: red curve.**

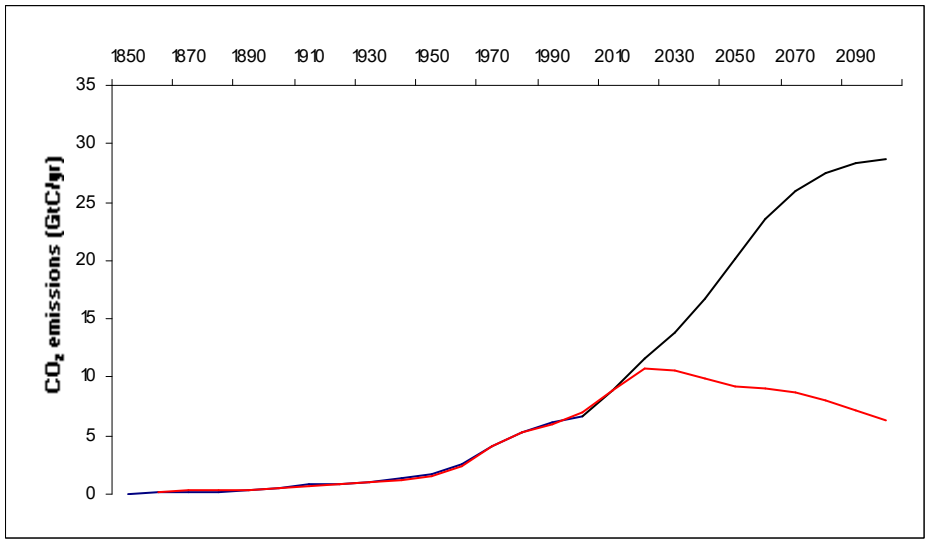

**Figure 11. $CO_2$ emissions trajectories (GtC/yr): RCP8.5 in comparison to peak fossil fuel scenario.**

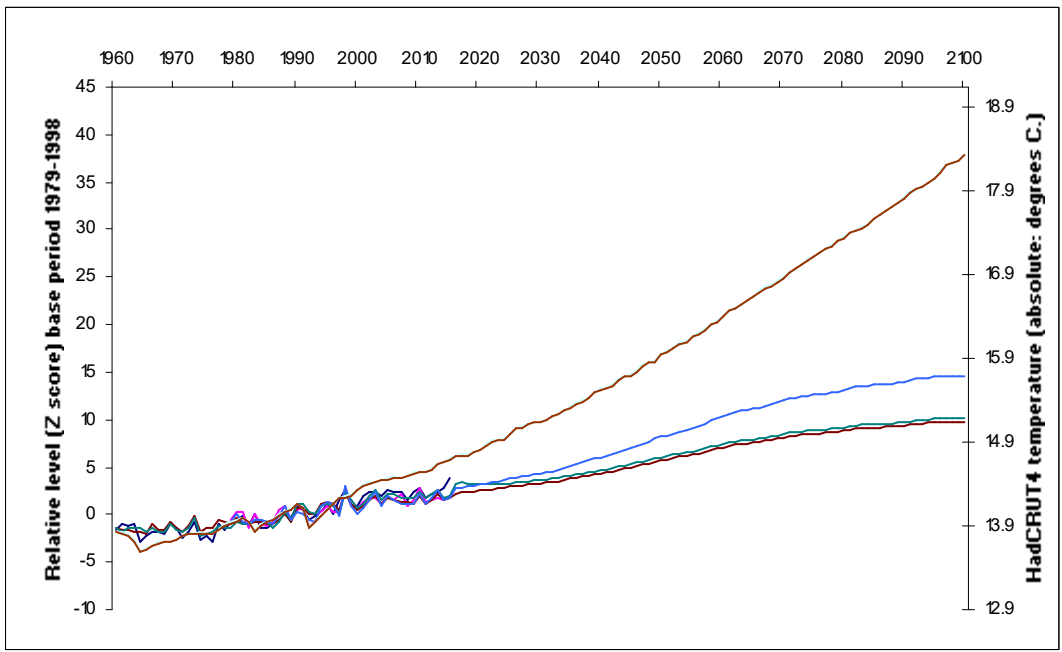

**Figure 12. Average global temperature reported to 2015 then modelled and forecast from the first-difference RCP8.5 atmospheric $CO_2$ scenario to 2100. UAH OLS model (blue curve); HadCRUT4: OLS model (red curve); dynamic regression model (green curve); reported temperatures to 2015: UAH (purple curve) and HadCRUT4 (dark blue**

curve). All foregoing curves Z-scored. Also shown for comparison is the temperature from the base-case (undifferenced) RCP8.5 scenario: Z scored (light blue curve); and absolute temperature in degrees Celsius (brown curve).

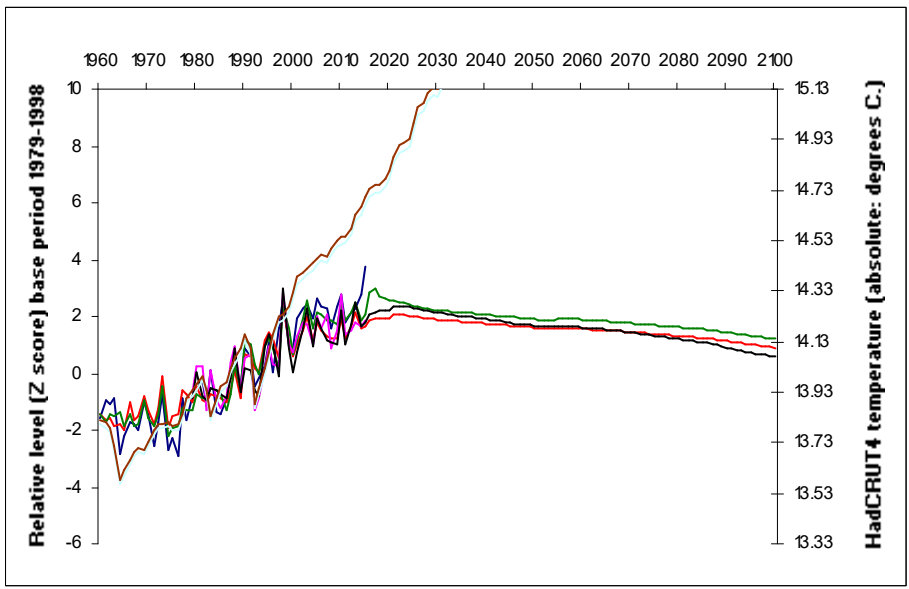

**Figure 13. Z scores. Temperature reported to 2015 then forecast for the first-difference peak fossil fuel atmospheric $CO_2$ scenario to 2100. UAH model (black curve); HadCRUT4: OLS model (red curve); dynamic regression model (green curve). Also shown for comparison is the temperature from the undifferenced RCP8.5 scenario: Z scored**

**(light blue curve); and in degrees Celsius (brown curve), and reported temperatures to 2015: UAH (purple curve) and HadCRUT4 (dark blue curve).**

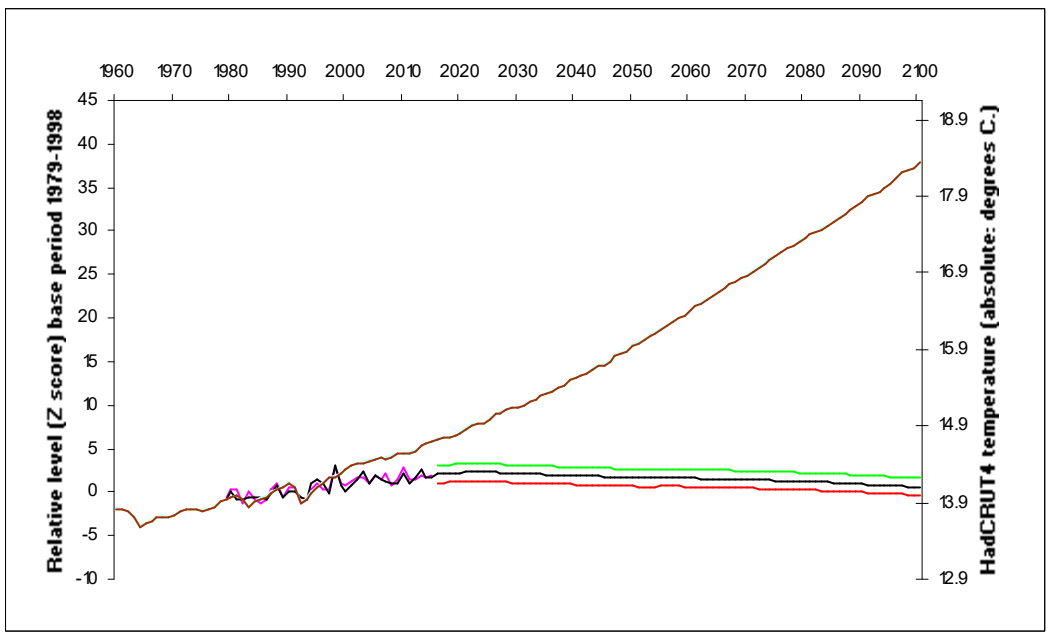

**Figure 14. Z scores. Forecast for UAH temperature for the first-difference peak fossil fuel atmospheric $CO_2$ scenario to 2100 (black curve) shown with 95% error bars and compared with the RCP8.5 temperature projection (brown curve). Also shown for comparison is the temperature from the undifferenced RCP8.5 scenario: Z scored (light blue**

curve); and in degrees Celsius (brown curve), and reported temperatures to 2015: UAH (purple curve) and HadCRUT4 (dark blue curve).

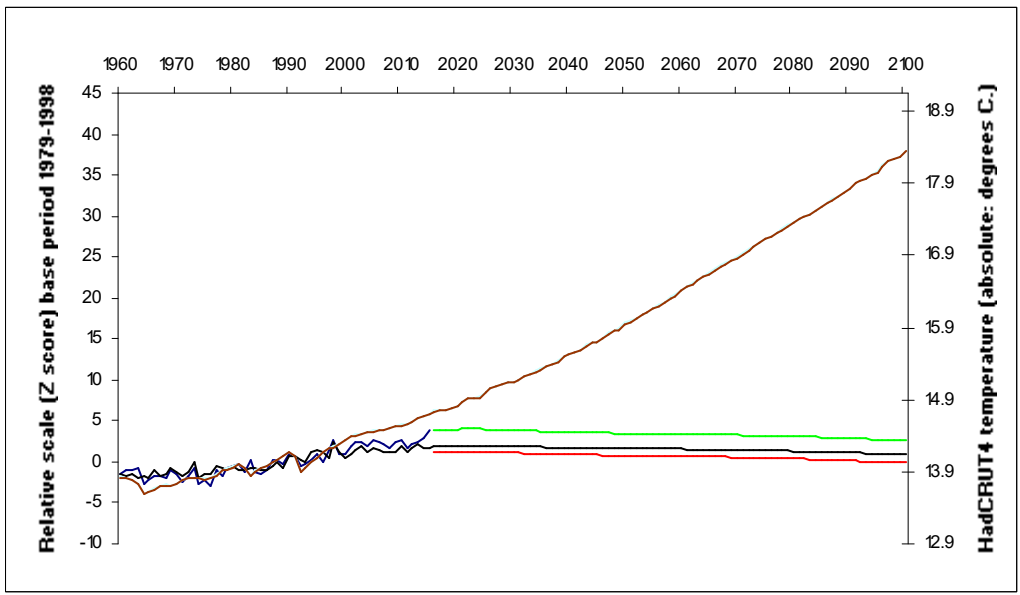

**Figure 15. Z scores. Forecast for HadCRUT4 temperature for the first-difference peak fossil fuel atmospheric $CO_2$ scenario to 2100 (black curve) shown with 95% error bars and compared with the RCP8.5 temperature projection. Also shown for comparison is the temperature from the undifferenced RCP8.5 scenario: Z scored (light blue curve);**

and in degrees Celsius (brown curve), and reported temperatures to 2015: UAH (purple curve) and HadCRUT4 (dark blue curve).